\documentclass[11pt,letterpaper]{article}

\usepackage{authblk}
\usepackage{xcolor}
\usepackage{array}
\usepackage[title]{appendix}
\usepackage{graphicx}
\usepackage{geometry}

\title{\Large What we talk about when we talk \\ about physics problem solving}
\author[1]{\normalsize Noa Perlmutter}
\author[2]{Zosia Krusberg}
\affil[1]{Program in Cognitive Science, The University of Chicago}
\affil[2]{Department of Physics, The University of Chicago}

\date{}

\begin{document}
\maketitle




\section*{Introduction}

I am a second-year cognitive science major, and as a student who has completed my physical science distribution requirements, I will likely never again encounter Gauss’s law. So why do I feel that the time and effort I devoted to solving Gauss’s law problems was worth it? Partly, I simply enjoy learning and the new perspective on the physical world that comes from understanding electromagnetism. But I was also fascinated by how physics problems train the mind in effective problem-solving strategies---of course I was, being a cognitive science major!

Two themes emerged as I reflected on this realization. First, physics problems serve as toy models for more complex problems outside of physics, cultivating broadly transferable problem-solving skills. Second, the process of solving physics problems invites reflection on our own cognitive and affective processes. These themes are deeply interconnected. A richer metacognitive understanding of our minds enables us to tackle more complex problems, while engaging with increasingly challenging problems, in turn, deepens our self-understanding.

In what follows, Professor Zosia Krusberg and I consider nine general lessons offered by the physics problem-solving process.





\section*{Lesson One: Confidence}

Physics problems often strike us as impossible at first, provoking stress and self-doubt. We struggle to make sense of the wording, visualize the system, distinguish relevant from irrelevant details, and identify which physical principles apply. Yet after working through countless problems that initially seemed overwhelming, we begin to develop confidence in our ability to find a way forward---both within and beyond the physics classroom. When we later encounter real-world problems that feel similarly daunting, {\bf physics problem solving reminds us not to let first impressions discourage us.} After all, few mental states are less useful to a problem solver than feeling overwhelmed: it shuts down focus and curiosity, disabling two of our most powerful tools for making sense of the world.\footnote{A wonderful example of what can be accomplished when someone is not overwhelmed by a problem is the story of the American mathematician George Dantzig. As a graduate student, Dantzig once arrived late to a statistics lecture, and assumed that the two problems written on the board were homework problems. When he handed in solutions to the problems a few days later, his professor informed him that he had just solved two previously unsolved statistics problems.} 


\section*{Lesson Two: Clarity}

Once we have convinced ourselves that a problem is solvable, we must decide how to begin. In physics, a few simple steps help launch the process: draw a clear diagram of the system, label the relevant variables, and identify the goal of the problem. These techniques work because {\bf successful problem solving begins with a clear and concrete picture of the situation at hand.} 

The same approach applies beyond physics. We often describe complex real-world issues broadly---``the problem of climate change'' or ``the problem of educational inequity''—yet to make progress, we must first clarify which specific aspect of the problem we are addressing. Only then can we focus our awareness, direct our attention, and begin to act effectively.


\section*{Lesson Three: Assumptions}

Physics problems also teach us that {\bf problem solving can be simplified through the productive use of assumptions} \cite{Fortus2009,Verostek2022}. In physics, recognizing the assumptions built into a problem---whether stated explicitly or implied---and invoking additional ones to simplify and constrain the situation are essential strategies. I developed the habit of stating my assumptions early in the process: noting, for example, whether dissipative forces could be neglected or whether nearby charges might affect the system. Doing so made it much easier to determine which physical principles to apply. 

Our minds likewise rely on both conscious and unconscious assumptions to simplify experience. Yet when addressing real-world problems, it is just as important to stay aware of these assumptions, to question them, and to be willing to let them go.


\section*{Lesson Four: Approximation}

While physics problem solving relies on assumptions and simplifications, the use of orders of approximation helps us identify the most important factors contributing to a solution and recognize when we may have gone too far in simplifying.

For example, to determine the trajectory of a basketball launched toward a hoop, a first-order solution might treat the ball as a point particle moving through a vacuum. A second-order correction could include air resistance and the ball’s rotation. Only if we sought extreme precision would we consider relativistic effects. In this way, {\bf orders of approximation simplify complex problems by providing a structured way to prioritize the most significant factors.}

This same technique also helps reveal when our simplifications fail. If an assumption neglects a factor whose contribution turns out to be an order of magnitude larger than the ones we included, something has clearly gone awry.


\clearpage

\section*{Lesson Five: Principles}

Once we have made sense of a problem, articulated our assumptions, and accounted for the most significant factors, we must identify the fundamental physical principles at play. One challenge in this step is that different physics problems may look similar on the surface yet require entirely different underlying physics—and the reverse can also be true.

For example, two collision problems might appear nearly identical, but in one, mechanical energy is conserved; in another, non-conservative forces must be included; and in yet another, relativistic effects become important. In each case, a different physical law governs the solution. {\bf Developing the ability to distinguish between fundamental principles and surface features is therefore a critical aspect of the problem-solving process} \cite{Chi1981,Mason2011,Chi2012}.

This skill also transfers beyond physics. Patients with the sniffles may all appear similar at first, but effective treatment depends on identifying the underlying cause---whether allergies, a bacterial infection, or the flu.


\section*{Lesson Six: Heuristics}

{\bf To move forward in problem solving, we rely on heuristics---cognitive tools and techniques built through experience} \cite{Polya1945,Schoenfeld1985}. In physics, effective heuristics include drawing new diagrams, comparing a problem to one we have solved before, or explaining it aloud---a technique I found particularly helpful.

Some heuristics also engage the unconscious mind, prompting insights that arise only after stepping away: brainstorming related concepts, taking a break, or even sleeping on the problem. From both a cognitive and practical perspective, it is fascinating that intense, conscious focus alone is sometimes insufficient; insights from the unconscious can play a vital role in the problem-solving process \cite{Hadamard1945,Spitz1993}.

Our problem-solving toolbox therefore includes both domain-specific and general heuristics that have proven useful through our unique experiences as problem solvers.


\section*{Lesson Seven: Collaboration}

Although our focus thus far has been on the inner, individual experience of problem solving, {\bf the diverse perspectives of others profoundly enrich the process}  \cite{Cox1993,Nemeth1995,McLeod1996,Guri2002,Guimera2005}. The study groups that formed in my physics classes often revealed the limits of my own understanding and the value of others’ cognitive approaches. When the groups worked well, there was something almost magical about how we complemented one another---the part of the problem that stumped me was exactly what someone else understood, and vice versa.

These groups also exposed the inherent challenges of collaboration: keeping an open mind to different perspectives, making space for every voice, and navigating disagreements with respect. Such dynamics exist on larger scales beyond the classroom. Science itself is a deeply collaborative enterprise. In research groups around the world, individuals come together to share reasoning, listen critically, exchange feedback, and use that feedback to deepen understanding. When obstacles arise, we turn to one another, since modern scientific problems demand combinations of insight and expertise that no single person can possess alone.

\section*{Lesson Eight: Reflection}

Once we have completed a problem, {\bf our solutions offer a valuable opportunity for reflection and introspection on the problem-solving process itself} \cite{Davidson1994,Mason2016}. At this stage, we can look back on the path from problem statement to solution, noting which strategies succeeded and which did not. We can also reflect on moments when we felt stuck and later experienced a sudden flash of insight---evidence of the unconscious mind at work---and consider what may have preceded that insight. Was it a new way of engaging with the problem? Something in how another person described it? Or simply the passage of time that allowed our minds to integrate information below the surface?

This is also the time to check whether our solutions make intuitive and mathematical sense, ensuring that every procedural step aligns with the outcome. A nonsensical answer, after all, signals that somewhere along the way an error has occurred.

Beyond these cognitive checks, we can reflect on our affective experience as well \cite{Debellis2006}. What emotions arose when we first encountered the problem, when we hit a dead end, or when understanding finally dawned? Such reflection prepares us for future problem-solving endeavors. It becomes easier to accept frustration and uncertainty when we recognize them as precursors to insight, joy, and curiosity---states that renew our motivation to engage with the next challenge.


\section*{Lesson Nine: Insight}

In the same way, {\bf mistakes offer valuable insight into our cognitive patterns and invite us to construct new and more effective ones} \cite{Fosnot1996}. By deliberately reflecting on feedback from graded assignments, we can identify recurring tendencies---misinterpreting problem statements, omitting essential details from diagrams, or performing incorrect mathematical operations. Once we recognize these patterns, we can create internal cues or external reminders to alert us before repeating them, gradually becoming more effective problem solvers.

Even errors that seem nonsensical at first can reveal how our minds are attempting to integrate new concepts or strategies \cite{Mason2016a}. Simply dismissing them---``That was silly; I’ll never do that again''---is rarely helpful and almost guarantees repetition. If, however, we take our mistakes seriously, they offer a remarkable opportunity to examine and refine the dynamic processes of our own thinking.\footnote{For an interesting study on what happens when students don’t reflect on their errors, see Andrew Mason and Chandralekha Singh. Do advanced physics students learn from their mistakes without explicit intervention? {\it American Journal of Physics}, 78(7):760, 2010.} If, on the other hand, we take the errors seriously, we are offered a wonderful opportunity to examine and modify the dynamic processes in our minds.




\section*{Conclusion}

We have argued that training in physics problem solving prepares us for the more complex challenges we encounter in life. Physics problem solving teaches us to approach uncertainty with confidence and to develop a clear understanding of both a problem and its context (Lessons 1--4). Few other domains provide such a structured way to cultivate a holistic understanding of a situation. The problems we tackle in physics are also unique in their balance of difficulty---demanding enough to require sophisticated, transferable strategies, yet simple enough to be fully solvable (Lessons 5--7). Finally, the immediate feedback inherent in physics coursework offers valuable opportunities to reflect on our problem-solving skills and on the cognitive and affective dimensions of our own learning (Lessons 8--9).

We therefore encourage physics instructors at all levels to help students recognize that physics problem solving is both complex and teachable, and to provide explicit, formal instruction in it. We also invite students to see their introductory physics courses as about more than Newton’s laws, Maxwell’s equations, or the laws of thermodynamics---as opportunities to cultivate powerful habits of thought and to explore the beautiful complexities of their own minds.

\section*{Acknowledgements}

We would like to extend our gratitude to Elam Coalson, Kacy Gordon, and Ella Herz for their extensive and insightful feedback on early drafts of this paper.

\addcontentsline{toc}{section}{References}
\bibliographystyle{unsrt}
\bibliography{standard-bibliography}

\end{document}